\documentstyle[preprint,pre,aps]{revtex}

\def\hx{\hat{\bbox{x}}}
\def\Bx{\bbox{x}}
\def\BX{\bbox{X}}
\def\Hx{\hat{\bbox{x}}}
\def\BrX{\breve{\bbox{X}}}

\begin{document}
\draft

\title{The partition function versus boundary conditions and
confinement in the Yang--Mills theory}

\author{\framebox{N.A.~Sveshnikov}$\,$ 
\and E.G.~Timoshenko\thanks{Corresponding author.
Permanent address:
Department of Chemistry, University College Dublin, Belfield, Dublin 4, Ireland.
E-mail: Edward.Timoshenko@ucd.ie}}
\address{
Nuclear Physics Institute, Moscow State University, 
Moscow, 119899, Russia
}
\date{\today}
\maketitle

\begin{abstract}
We analyse dependence of the partition function on the boundary 
condition for the longitudinal component of the electric field 
strength in gauge field theories.
In a physical gauge the Gauss law constraint may be
resolved explicitly expressing this component via an integral of 
the physical transversal variables. 
In particular, we study quantum electrodynamics with an external charge 
and $SU(2)$ gluodynamics. 
We find that only a charge distribution slowly decreasing at spatial infinity
can produce a nontrivial dependence in the Abelian theory.
However, in gluodynamics for temperatures below some critical value
the partition function acquires a delta--function like dependence on
the boundary condition, which leads to colour confinement.
\end{abstract}


\centerline{*\ *\ *}

\begin{quotation}

It is with a feeling of deep sadness and irrecoverable loss
that I had to complete writing this paper on my own.
In the name of N.A. Sveshnikov Theoretical Physics has lost
an extremely talented scientist, a great educator and a very nice person.
His ideas and elegant scientific style had a big influence on his students,
one of which I was privileged to be, and on many colleagues who knew him well.
Perhaps, this paper may be viewed as a logical conclusion to years of
our joint research on the problem of confinement. 
It was one of N.A. Sveshnikov's gifts to be able
to see farther ahead in search for the hidden mathematical beauty of
the physical world, the beauty that expresses the essence of all things.

\rightline{E. T.}
\end{quotation}

%Sec1
\section{Introduction}

There is a considerable tradition in the field theory 
(see e.g. Ref. \cite{FadSlav}) to neglect any
surface terms inevitably appearing in derivations.
This is usually motivated by a fast decrease of all fields at 
spatial infinity. Such behaviour is only natural for 
theories with short--range interactions, 
but it is by no means obvious if long--range interactions come into play.

The same problem acquires a somewhat different form at finite temperature.
Obviously, the partition function of a translationally invariant system
is ill--defined in infinite volume. Therefore, initially one has 
to enclose the system into
a finite domain $V$ and to assume some boundary conditions at the
boundary $\partial V$.
Then, to analyse dependence on boundary conditions we have to consider
the functional,
\begin{equation}
Z[\chi] = \mbox{Tr} \left( e^{-\beta H_V}\,
\delta(\phi|_{\partial V}-\chi)\right),
\end{equation}
where $H_V$ is the Hamiltonian of the system in volume $V$, $\beta$ is the
inverse temperature, $\phi$ is some subset of the canonical variables
and function $\chi$ defined on the boundary $\partial V$ specifies
the Dirichlet boundary conditions for the latter.
In this framework we shall call $Z[\chi]$ the {\it effective partition
function}, its introduction being analogous in spirit 
to that of the effective action.
It is reasonable to expect that $Z[\chi]$, and any thermodynamic
function, become independent of the particular choice of boundary 
conditions in the thermodynamic limit $V\rightarrow\infty$.
Once again, one may doubt whether that would really be the case 
for long--range interacting systems.

It has been early recognised in the theory of gravity 
\cite{RegTet} that there are important physical
situations in which boundary terms may have a physical meaning.
We would like to mention also that many problems in hydrodynamics, 
such as e.\ g.\ description of 
surface waves \cite{Lew}, do require to consider variables at the boundary
and nonvanishing surface terms.

Recently the interest to boundary effects in various field theories
has been rapidly increased \cite{NewBou}.
It has been found in the framework of the algebraic quantum field theory 
\cite{MorStr} that certain 2-dimensional models
possess a nontrivial dynamics of the {\it variables at infinity}, and that such 
dynamics is responsible for the phenomenon of the dynamic mass generation.

Our current purpose is to emphasise the role of boundary terms
in the 4-dimensional gauge field theory at finite temperature and to
study physical effects they can produce.
In this case the appearance of a nontrivial $\chi$-dependence in $Z[\chi]$
in the thermodynamic limit is almost obvious for the following reasons.
The gauge theory is initially formulated in terms of an enlarged set
of variables, the vector potentials $\bbox{A}$ and the electric
field strengths $\bbox{E}$ in the Hamiltonian formulation,
that make the gauge invariance explicitly manifest.
Further, to obtain observable quantities 
one has to project the theory onto a subset of the 
{\it physical} variables by resolving the
Gauss law constraint, $\bbox{\nabla E}=\rho$, and by
adopting a gauge condition. Boundary conditions, of course, have to
be compatible with these. For instance, in the Abelian case consider 
the boundary condition for the electric field strength component $E_\Vert$
normal to the boundary $\partial V$, which we take for simplicity as
a sphere of radius $R$,
\begin{equation}
\label{Echi}
R^2\,E_\Vert(R\hx)\, |_{\partial V}=\chi(\hx).
\end{equation}
This variable then has to obey the integrated form of the Gauss law,
and therefore,
\begin{equation}
\label{Zzchi}
Z[\chi] \propto \delta(\int_{\partial V} d\hx\,\chi - \int_V d\bbox{x}\,\rho).
\end{equation}

We would like to emphasise that the dependence of the effective partition
function on the boundary condition imposed on $E_\Vert$ is of primary
importance in the gauge theory because of the direct relation of this
component to the colour charge flux due to the Gauss law. Analysis of namely
this dependence will be the subject of the current paper.
Based on the knowledge of this dependence alone we can suggest a simple
{\it confinement criterion}:
\begin{equation}
\label{ConfCrit}
Z[\chi]\propto \prod_{\hx} \delta(\chi(\hx)).
\end{equation}
The latter condition simply means that the colour flux 
is strictly zero in {\it every} direction at spatial infinity
for any state belonging to the Hilbert space of the system.

The plan of the paper is as follows. In Sec. 2 we proceed with a careful
calculation of of $Z[\chi]$ in the simplest case of the Abelian theory
with an external charge density. Then, in Sec. 3 we reproduce the answer
obtained in the previous section using another technique, which is also
applicable to the non--Abelian theory. Sec. 4 is devoted to calculation
of $Z[\chi]$ using the mean--field approximation and to consequent analysis
of the confinement phase transition in gluodynamics.

\section{Partition Function of QED with an External Charge}

The partition function in a finite domain $V$
may be represented by an Euclidean path integral over the fields  periodic
in time on the interval $[0,\beta]$, where $\beta$ is the inverse temperature.
The path integral is well--defined only if some
boundary conditions are specified on all fields at the boundary $\partial V$.

Let us denote the transversal and longitudinal components of vectors 
with respect to the gradient $\partial$ 
as superscripts $\bbox{\partial} \bbox{A}^{\perp} = 0$, and with respect to
the vector $\bbox{x}$ as subscripts $\bbox{x\, A}_{\perp} =0$.
For simplicity we choose the Coulomb gauge, $\bbox{A} = \bbox{A}^{\perp}$.
We note also that the condition 
$\bbox{A} = \bbox{A}_{\perp}$ corresponds to the Fock--Schwinger gauge 
\cite{FockSch} (see Appendix A).
We shall also assume that $V = \{ \bbox{x}:\ |\bbox{x}| = R\}$ is a spherical
domain with the radius $R$.
The Gauss law constraint,
\begin{equation}
\bbox{\partial} \bbox{E} = \rho(\bbox{x}) \label{GauLaw}
\end{equation}
allows to eliminate one space component of the electric field strength 
$\bbox{E}$.

In a previous work \cite{NewProc} on the basis of the general result \cite{VOS}
we have developed the Hamiltonian formalism
for the system in a finite spherical domain incorporating the boundary values
as Hamiltonian variables.
We have shown that the boundary conditions 
$\bbox{E}_{\perp}(R\hx) =0$ and $x_{j}F_{ij}(R\hx)=0$
are consistent with the localised time evolution in the Fock--Schwinger gauge.
By transforming the theory to the Coulomb gauge 
one would arrive instead at the boundary condition of the form
$\bbox{E}^{\perp}(R\hx) =0$.
Since these variables are independent,
the dependence on a particular choice of the boundary conditions
disappears for infinite system. 
The situation is quite different for the component 
$E_{\Vert}(R\hx) = \hx \bbox{E}(R\hx)$. Indeed,
Eq. (\ref{GauLaw}) may be easily solved,
\begin{equation}
E_{\Vert}(\bbox{x}) = \frac{1}{x^{2}}\int_{0}^{x}y^{2}dy\, \bigl(\rho -
\bbox{\partial} \bbox{E}_{\perp}\bigr)(y\hx)\,.
\label{SolGauLaw}
\end{equation}
Both types of the transversal variables are connected by the relation,
\cite{SveTim}
\begin{equation}
\bbox{E}_{\perp}(\bbox{x}) = \bbox{E}^{\perp}(\bbox{x}) 
- \bbox{\partial} \int_{0}
^{x} dy\,\hx\bbox{E}^{\perp}(y\hx)\,.
\label{K}
\end{equation}
Combination of Eqs. (\ref{SolGauLaw}) and (\ref{K}) now yields,
\begin{equation}
R^{2} E_{\Vert}(R\hx)- \int_{0}^{R}y^{2}dy\,\rho(y\hx) =
\hat{\Delta} \int_{0}^{R} (R-y)dy\,\hx\bbox{E}^{\perp}(y\hx)\,,
\label{BouRel}
\end{equation}
where we have used the spherical part of the Laplacian
$\Delta \equiv x^{-1}\partial^{2}_{x}\,x + x^{-2}\hat{\Delta}$.
It is clear that the requirement $E_{\Vert}(R\hx)=\chi(\hx)$,
where $\chi(\hx)$ is arbitrary, is nothing but a
constraint on the physical variables $\bbox{E}^\perp$. 
As we have seen \cite{NewProc}, it is this constraint
that makes the finite volume Hamiltonian formalism closed.

Now then,
the partition function of QED with an external charge may be represented by
the following path integral in the Coulomb gauge,
\begin{eqnarray}
&& Z = \int {\cal D}\bbox{A} {\cal D}\bbox{E}\ 
\delta(\bbox{\partial} \bbox{A})\,
\delta(R^{2} E_{\Vert}(R\hat{\bbox{x}}) - \chi(\hat{\bbox{x}})) 
\nonumber\\ &&
\exp \int_{\Lambda}d^{4}x\left(i\,\bbox{E}\dot{\bbox{A}} - \frac{1}{2}\bbox{E}
^{2} + \frac{1}{2}\bbox{A}\Delta\bbox{A} 
- \frac{1}{2\epsilon^{2}}(\bbox{\partial}
\bbox{E} - \rho)^{2}\right)\,, \label{Z}
\end{eqnarray}
where we have used the notation for the domain $\Lambda=[0,\beta] \times V$
(we shall also use the notation $\partial\Lambda=[0,\beta]\times\partial V$).
We have also introduced a regularisation of the Gauss law by
the parameter $\epsilon$, which should be set equal to zero at the end of 
calculations.
This Gaussian integral is evaluated in a standard manner by a shift 
of the integration variables.
To find the integral over $\bbox{E}$ we introduce a new
integration variable, $\bbox{E}_{1}$,
\begin{equation}
\bbox{E} = \bbox{E}_{1} + \bbox{{\cal E}}\,, 
\qquad \bbox{E}_{1}(R\hat{\bbox{x}}) 
= 0\,. \label{EEp}
\end{equation}
Here the new variable $\bbox{E}_{1}$ satisfies the zero boundary condition
and $\bbox{{\cal E}}$ is chosen so that there is no linear term 
in $\bbox{E}_{1}$.
This gives the equation on $\bbox{{\cal E}}$,
\begin{eqnarray}
&&i\dot{\bbox{A}} - \bbox{{\cal E}} + 
\frac{1}{\epsilon^{2}}\bbox{\partial}(\bbox{\partial}\bbox{{\cal E}})
- \frac{1}{\epsilon^{2}}\bbox{\partial} \rho = 0\,, 
\label{eqE} \\
&& R^{2}\hat{\bbox{x}}\bbox{{\cal E}}(R\hat{\bbox{x}}) 
= \chi(\hat{\bbox{x}}) \,.
\label{boE}
\end{eqnarray}
The latter boundary condition follows from the second delta--function
in Eq. (\ref{Z}).
We may decompose this vector onto the transversal and longitudinal parts in 
the momentum space, $\bbox{{\cal E}} = \bbox{{\cal E}}^{\perp} 
- \bbox{\partial}\varphi$.
Then, the transversal part is simply $\bbox{{\cal E}}^{\perp} = i\dot{\bbox{A}}$,
and the equation for $\varphi$ becomes,
\begin{eqnarray}
&& (\Delta - \epsilon^{2}) \varphi = -\rho\,, \label{eqFi} \\
&& R^{2}\frac{\partial\varphi}{\partial R} = - \chi(\hat{\bbox{x}})\,.
\label{boFi}
\end{eqnarray}
The partition function (\ref{Z}) is further
decomposed as the product,
\begin{eqnarray}
&&  Z = Z_{1} \tilde{Z}\,, \nonumber\\
&&  Z_{1}= \int {\cal D}\bbox{A}^{\perp}{\cal D}\bbox{E}
\,\exp\int_{\Lambda} d^{4}x \,
\left(-\frac{1}{2}\dot{\bbox{A}}^{2} +
\frac{1}{2}\bbox{A}\Delta\bbox{A} - \frac{1}{2}\bbox{E}_{1}^{2}
- \frac{1}{2\epsilon^{2}}(\bbox{\partial} \bbox{E}_{1})^{2} \right) \,, 
\label{Zp} \\
&& \tilde{Z} = \exp \beta \left( \frac{1}{2}\int
_{\partial V} d\hat{\bbox{x}}\,
\chi(\hat{\bbox{x}}) \varphi(R\hat{\bbox{x}}) - \frac{1}{2}\int_{\partial V}
d\bbox{x}\,\rho(\bbox{x})\varphi(\bbox{x}) \right)\,, \label{Zt}
\end{eqnarray}
where $\varphi$ is the solution of Eqs. (\ref{eqFi},\ref{boFi}).

The solution of Eqs. (\ref{eqFi},\ref{boFi}) is, obviously, the sum of
the homogeneous part, $\phi$,  satisfying  the nontrivial boundary 
condition, and of
the inhomogeneous part satisfying the zero boundary condition,
\begin{equation}
\varphi = \phi - G \bullet \rho\, \qquad G = (\Delta - \epsilon^{2})^{-1}\,,
\label{FiFi}
\end{equation}
where $G$ is the
Green function corresponding to the zero Neumann boundary condition at $R$.
The effective partition function can be presented as,
\begin{eqnarray}
\tilde{Z} &=& \tilde{Z}_{\chi}\tilde{Z}_{\rho\rho}\tilde{Z}_{\rho\chi}\,, 
\nonumber \\
\tilde{Z}_{\chi} &=& 
\exp \left( \frac{\beta}{2}\int_{\partial V} d\hat{\bbox{x}}\,
\chi(\hat{\bbox{x}}) \phi(\hat{\bbox{x}}) \right)\,, \label{Zchi} \\
\tilde{Z}_{\rho\rho} &=& \exp \left(\frac{\beta}{2}
\int_{\partial V} d\bbox{x}d\bbox{y}\,
\rho(\bbox{x}) G(\bbox{x},\bbox{y})\rho(\bbox{y})\right)\,, \label{Zrho} \\
\tilde{Z}_{\rho\chi} &=& 
\exp\left(- \frac{\beta}{2}\int_{\partial V} d\bbox{x}\, 
\phi(\bbox{x}) \rho(\bbox{x}) \right) 
\exp \left(-\frac{\beta}{2} \int_{\partial V} d\hat{\bbox{x}}\,
\chi(\hat{\bbox{x}})\,(G\bullet\rho)(\hat{\bbox{x}})\right). \label{Zrhochi}
\end{eqnarray}
It is natural to consider the problem further
in terms of the spherical coordinates.
Solution of Eq. (\ref{eqFi}) regular inside the sphere is given by,
\begin{eqnarray}
&& \phi_{lm} = C_{lm}\,\sqrt{\frac{\pi}{2\epsilon r}}I_{l+1/2}(\epsilon r)\,, 
\label{Film} \\
&& I_{n-1/2}(z) = \sqrt{\frac{2}{\pi z}} z^{n} \left(\frac{1}{z}\frac{d}{dz}
\right)^{n}\,\cosh z\,.
\end{eqnarray}
The constant $C_{lm}$ is determined from Eq. (\ref{boFi}).
The role of the regulator $\epsilon$ now becomes clear. The zero mode
solution is simply, 
\begin{equation}
\phi_{00} = C_{00} \frac{\sinh \epsilon r}{\epsilon r}\,,
\qquad C_{00} = -\frac{\chi_{00}}{R^{2}\epsilon (\cosh \epsilon R/
\epsilon R - \sinh \epsilon R/(\epsilon R)^{2})}\,,
\label{Fioo}
\end{equation}
and it is $1/\epsilon^{2}$ singular as $\epsilon$ tends to zero.
At the same time, solutions for other modes are perfectly regular in
this limit and tend to,
\begin{equation}
\phi_{lm} = C_{lm}\,r^{l}\,,\qquad C_{lm} = -\frac{\chi_{lm}}{lR^{l+1}}\,.
\end{equation}
In consideration of the zero mode one must therefore be more careful and keep 
$\epsilon$ nonvanishing.
The zero mode Green function, that is defined by,
\begin{eqnarray}
&&\left(\frac{1}{r}\frac{\partial^{2}}{\partial r^{2}}r - \epsilon^{2}
\right) G_{00}(r,r') = \frac{\delta(r-r')}{rr'}\,, \\
&&\frac{\partial G_{00}(r,r')}{\partial r}\biggl.\biggr|_{r = R} = 0\,,
\end{eqnarray}
is easily calculated,
\begin{eqnarray}
 G_{00}(r,r') &=& \frac{1}{\epsilon r r'}\biggl(\biggr. 
\frac{1}{2} \sinh \epsilon
|r-r'| - \frac{1}{2}\sinh \epsilon(r+r') \nonumber\\
&+& \frac{\sinh \epsilon R - \cosh \epsilon R/\epsilon R}
{\cosh\epsilon R - \sinh \epsilon R/\epsilon R}
\sinh \epsilon r \sinh \epsilon r' \biggl.\biggr)\,.
\end{eqnarray}
The leading terms at small $\epsilon$ are,
\begin{eqnarray}
 G_{00}(r,r') &\simeq& -\frac{3}{\epsilon^{2}R^{3}} + \frac{9}{5 R}
- \frac{1}{2} \frac{r^{2} + r'^{2}}{R^{3}} -\frac{1}{\max (r,r')}\,, \\
 \phi_{00} &\simeq&- \left(\frac{3}{\epsilon^{2}R^{3}} - \frac{3}{10 R}
+ \frac{r^{2}}{2 R^{3}}\right) \chi_{00}\,.
\end{eqnarray}
Substitution of these results into formulae (\ref{Zchi}-\ref{Zrhochi})
gives for the nonzero modes in the limit $\epsilon = 0$,
\begin{equation}
\tilde{Z}_{\chi\,l>0} = 
\exp\left(-\frac{\beta}{2R}\sum_{lm > 0} \frac{|\chi_{lm}|^{2}}{l}\right)\,.
\end{equation}
For simplicity, we assume that the distribution of the charge $\rho$ is
spherically symmetrical. In this case only the zero mode term survives.
If we introduce the charge density momenta,
\begin{equation}
{\cal Q}_{00} = \int_{0}^{R} r^{2}\,dr\,\rho_{00}(r)\,, \qquad
{\cal G}_{00} = \int_{0}^{R} r^{4}\,dr\,\rho_{00}(r)\,,
\end{equation}
our results may be summarised as follows,
\begin{eqnarray}
\tilde{Z}_{00} &=& \exp \beta\biggl(\biggr.
-\frac{1}{2}\int_{0}^{R} x^{2}dx\,y^{2}dy\,\frac{\rho_{00}(x)\,
\rho_{00}(y)}{\max(x,y)}
-\frac{3}{2\epsilon^{2}R^{3}}
({\cal Q}_{00}-\chi_{00})^{2} \nonumber\\
&+& \frac{1}{R}(\frac{9}{10}{\cal Q}_{00}^{2}
-\frac{1}{10}\chi_{00}^{2} - \frac{3}{10}\chi_{00}{\cal Q}_{00})
-\frac{1}{2R^{3}}({\cal Q}_{00}-\chi_{00}){\cal G}_{00} \biggl.\biggr)\,.
\end{eqnarray}
In the limit $\epsilon \rightarrow 0$ this 
functional contains the delta--function 
of the condition ${\cal Q}_{00} = \chi_{00}$,
and in addition we find the following correction to the standard answer
due to the surface terms,
\begin{equation}
\tilde{Z}_{00} = \exp\left(\frac{\beta}{2R}{\cal Q}_{00}^{2}\right)\,
\delta({\cal Q}_{00} - \chi_{00})\,.
\end{equation}

As a simple illustration of the above result it is instructive 
to consider the 
charge density $\rho_{00} = - \kappa/(\sqrt{\pi}\,r)$  corresponding to
a linear rising electric potential $\varphi = \kappa\,r$.
For such an exotic charge distribution, 
modeling the confinement alike potential,
we find the additional constant contribution to the free energy density,
\begin{equation}
\Delta {\cal F} = - \frac{\log Z}{\beta V} = -\frac{3}{32\pi^{2}}
\kappa^{2}\,.
\end{equation}
It is interesting to note that this correction
makes the free energy density smaller, and in this sense the boundary
effects are thermodynamically significant.
This example exhibits a promising connection between 
the boundary effects and the confinement phenomenon.

%Sec2
\section{Collective variable formulation}

In the present section we give a different formulation of the same
problem by introducing the collective variable, $\sigma$, conjugate 
to the Gauss law constraint. Both formulations are completely 
equivalent in the Abelian theory,
the transformation between them being just a trivial change of variables.
However, the new formulation appears to be more fruitful in the non--Abelian
gauge theory. Let us rewrite formula (\ref{Z}) in terms of the collective 
variable $\sigma$, introduced by the definition,
\begin{equation}
\exp\left(-\frac{1}{2\epsilon^{2}}
\int_{\Lambda}d^{4}x\,(\bbox{\partial} \bbox{E}
-\rho)^{2} \right) = \int {\cal D}\sigma\,\exp\int_{\Lambda}d^{4}x\,
\left(-\frac{\epsilon^{2}}{2}\sigma^{2} 
+ i\sigma(\bbox{\partial} \bbox{E} -\rho)\right).
\end{equation}
One starts by taking the integral over $\bbox{E}$,
\begin{eqnarray}
&& J= \int {\cal D}\bbox{E}\, \exp\int_{\Lambda}d^{4}x\, \left(
-\frac{1}{2}\bbox{E}^{2} +
i\bbox{E}\dot{\bbox{A}}+ i\sigma\bbox{\partial}\bbox{E}\right)\,\delta(R^{2}
E_{\Vert}(R\hx)-\chi(\hx)).\  
\end{eqnarray}
This can be done in analogy with the previous 
section by applying a shift $\bbox{E}=
\bbox{E}_{1}+\bbox{{\cal E}}$ and using the decomposition
$\bbox{{\cal E}}=\bbox{{\cal E}}^{\perp}-\bbox{\partial}\varphi$.
Then, we shall get: $\bbox{{\cal E}}^{\perp}=i\dot{\bbox{A}}$ 
and $\varphi = i\sigma$.
Using the delta--function  in the above formula one
may rewrite the boundary term as 
$i\int_{\partial \Lambda} dt\, d\hx\,\chi \sigma$. 
In the term $\int d\bbox{x}\,\dot{\bbox{A}}\bbox{\partial} \sigma$
the appropriate boundary term vanishes by the gauge condition.
Thus, we obtain,
\begin{eqnarray}
&& J = \exp\left(-\frac{1}{2}\int_{\Lambda} 
d^{4} x(\dot{\bbox{A}}^{2} 
+(\bbox{\partial}\sigma)^{2}) + 
i\int_{\partial \Lambda} dt\, d\hx\,\chi \sigma\right)\,
\delta(iR^{2}\frac{\partial \sigma}{\partial R}+\chi)
.\ 
\end{eqnarray}
At this stage the $\chi$ dependence is contained only in the integral,
\begin{eqnarray}
\tilde{Z} &=& \int{\cal D}\sigma\,\exp\left(-\int_{\Lambda} d^{4}x\,\bigl(
\frac{1}{2}(\bbox{\partial} \sigma)^{2}
+\frac{\epsilon^{2}}{2}\sigma^{2} +i\sigma\rho \bigr)
\right.
\nonumber\\ 
&+& \left. i\int_{\partial\Lambda} dt\, d\hx\,\chi \sigma\right)
\delta(iR^{2}\frac{\partial \sigma}{\partial R}+\chi)
\,.
\end{eqnarray}
Such integral is evaluated
straightforwardly by a shift of the integration variable, $\sigma =\sigma_{1}
+\varsigma$. The new variable satisfies the trivial boundary condition,
$R^{2}\partial\sigma_{1}(R\hx)/\partial R = 0$, 
and $\varsigma$ is found by the requirement 
that there is no term linear in $\sigma_{1}$ after the shift. 
This gives the equations on $\varsigma$,
\begin{equation}
(\Delta-\epsilon^{2})\varsigma = i\rho\,, \quad
R^{2}\frac{\partial \varsigma}{\partial R}= i\chi.
\end{equation}
Using the boundary condition on $\varsigma$ we finally derive,
\begin{equation}
\tilde{Z} =\exp\left( 
-\frac{i}{2} \beta \int_{V} d\bbox{x}\, \varsigma\rho
+\frac{i}{2}\beta\int_{V} d\hx\,\varsigma\,\chi
\right)\,.
\end{equation}
Note that with the
identification $\varsigma = -i\varphi$ we have precisely 
reproduced the answer of Eqs. (\ref{eqFi},\ref{boFi},\ref{Zt}).

%Sec3
\section{Non--Abelian theory}

In a previous paper \cite{SveTim} we have derived a representation for the
partition function $Z$ of SU(N) gluodynamics in a finite volume
as the path integral over the collective
variables analogous to that of the previous section.   
In order to find the effective partition function
$Z[\chi]$ dependence we have to return to the beginning
of that derivation in the Fock--Schwinger gauge,
\begin{eqnarray}
Z[\chi] & = & \int {\cal D}\bbox{A}\,{\cal D}\bbox{E}\,{\cal D}\sigma\,
\exp\biggl( \int_\Lambda d^4 x \bigl(
i\,\bbox{E}\dot{\bbox{A}}-\frac{1}{2}\bbox{E}^2-\frac{1}{2}\bbox{B}^2 
\bigr.\biggr. \nonumber\\
 & + & i\,\sigma\bbox{\nabla}\bbox{E} \biggl.\bigl.\bigr)\biggr)
\, \delta(A_\Vert)\,\delta(R^2\,E_\Vert(R\hx)-\chi),
\end{eqnarray}
where $\sigma$, clearly, is just a different notation for
the temporal component of the gauge field, $A_0$. Obviously,
any dependence on $\chi$
is concentrated in the path integral over $E_\Vert$,
\begin{equation}
\label{intEVert}
I=\int {\cal D}E_\Vert \exp \biggl( \int_\Lambda d^4 x \bigl(
-\frac{1}{2}E_\Vert^2+i\,\sigma\,(\bbox{\partial}\hx)E_\Vert\bigr)\biggr)\,
\delta(R^2 E_\Vert-\chi).
\end{equation}
This is calculated by a shift $E_\Vert=E_\Vert^1+{\cal E}$, where 
${\cal E}=-i\,\partial \sigma/\partial x$ and $E_\Vert^1$ satisfies the zero
Dirichlet boundary condition. Such a derivation gives,
\begin{equation}
\label{Iii}
I=\exp \biggl( \int_0^\beta dt \biggl[- \frac{1}{2} \int_V d\bbox{x} 
\biggl( \frac{\partial \sigma}{\partial x}\biggr)^2
+i\int d\hx \sigma\,\chi \biggr]\biggr)\,
\delta(R^2 \frac{\partial \sigma}{\partial R}-i\chi).
\end{equation}
Next, by introducing the path integral representation,
\begin{equation} \label{NuEq}
\exp\left( -\frac{1}{2}\int_\Lambda d^4 x\, B_\Vert^2\right)
=\int {\cal D}\nu\,\exp\left( \int_\Lambda d^4 x\, \left(
-\frac{1}{2}\nu^2+i\,\nu\,B_\Vert\right)\right),
\end{equation}
and after performing the
integrations over $\bbox{A}_\perp$ and $\bbox{E}_\perp$ 
(see Ref. \cite{SveTim} for more details) we finally obtain,
\begin{eqnarray} 
 Z[\chi] &=& \int {\cal D}\sigma\,{\cal D}\nu\, \exp(-W[\sigma,\nu]
+i\int_{\partial\Lambda}dt\,d\hx\,\sigma\chi)
\,\delta(R^{2}\sigma'-i\chi)
\,, \nonumber \\
 2W[\sigma,\nu] &=& \nu\bullet\nu + \bbox{\partial} \sigma
\bullet \bbox{\partial} \sigma + \bbox{K}_{-}\bullet C_{+}^{-1}\bullet 
\bbox{K}_{+} \nonumber \\
&+& \bbox{K}_{+}\bullet C_{-}^{-1}\bullet \bbox{K}_{-} 
+ \mbox{tr}\log C_{+}C_{-}\,, \label{Zzci} \\
C_{\pm} &=& - \Delta_{x} - \nabla_{t}^{2} \pm D\,, \quad 
\bbox{K}_{\pm} =
\bbox{\partial}_{\pm}\nu \pm \nabla_{t}\bbox{\partial}_{\pm}\sigma\,, \nonumber\\
\nabla_{t}^{ab}&=&\delta^{ab}\partial_{t} - gt^{abc}\,\sigma^{c}\,, 
\quad D^{ab} = gt^{abc}\,\nu^{c}\,, \nonumber
\end{eqnarray}
where the projected derivatives are defined by,
\begin{equation}
\partial_{\pm}^{i}=\Pi^{ij}_{\pm}\,\partial_{j}\,,\quad
\Pi^{ij}_{\pm}=\frac{1}{2}(\delta^{ij}-\hat{x}^{i}\hat{x}^{j}\pm i
\epsilon^{ijk}\hat{x}^{k})\,
\end{equation}
and the bullet denotes the 4-d integration over the domain $\Lambda$.

In the saddle point approximation one expands the action near the saddle point,
\begin{eqnarray}
W[\varsigma + \sigma_{1}] &=& W[\varsigma]+ \int_{\Lambda} d^{4}x\,
\frac{\delta W}{\delta \varsigma(\bbox{x})} \sigma_{1}(\bbox{x})  \nonumber \\
&+& \int_{\partial\Lambda} dt\,R^{2}d\hx\,
{\cal E}^{(1)}[\varsigma]\,\sigma_{1}(R\hx)+
\ldots.
\end{eqnarray}
In the zero (mean--field) approximation we may write,
\begin{eqnarray}
\tilde{Z}[\chi] &=& \exp\left(-W[\varsigma]+i\int_{\partial\Lambda}
dt\,d\hx\, \varsigma \chi\right)\,, \\
\frac{\delta\,W}{\delta \varsigma} &=& 0\,, \qquad
R^{2}\frac{\partial \varsigma(R\hat{\bbox{x}})}{\partial R}
=i\chi,  \label{NonAbSys}
\end{eqnarray}
where the contribution from the  first Euler derivative of the action
${\cal E}^{(1)}[\varsigma]$ 
is precisely canceled with that from the surface term in Eq. (\ref{Zzci}).

Thus, in the mean--field approximation the dependence $Z[\chi]$ is controlled
by the saddle point solution $\varsigma$.
As we have seen in  the previous section, 
the Abelian gauge theory possesses only the trivial
solution $\varsigma=0$.

In Ref. \cite{SveTim} we have studied constant solutions
of the mean--field equations. 
Let us reproduce those results briefly here, but in addition
carefully keeping a finite volume.
For simplicity we also restrict ourselves to the gauge group $SU(2)$.
First of all, we can introduce the notations for the free energy density,
${\cal F}_{R}$,
\begin{equation}
W_{R} = \beta V_{R}\, {\cal F}_{R}\,, \quad 
{\cal F}_{R} = \gamma_{R}\,F_{R}\,, 
\quad \gamma_{R} = \frac{8\pi^{2} R \delta({\hat{0})}}
{\beta^{2}\,V_{R}}\,,
\end{equation}
where $V_{R} = 4\pi R^{3}/3$ is the domain volume and
$\delta(\hat{0})=\frac{1}{4\pi}\sum_{l}(2 l +1)$ is
the ultravioletly divergent angular delta--function with
coinciding arguments. The function $F_R$ is now expressed via
the dimensionless variables,
\begin{equation}
\sigma = \frac{2\pi \, s}{\beta \,g},\qquad
\nu = i \left(\frac{2\pi \,u}{\beta\,g}\right)^2\,,
\end{equation}
where to produce a real mean magnetic field $\nu$ has to be purely
imaginary (see Eq. (\ref{NuEq})).
After introducing the control parameter 
$a=(2\pi)^{4}/(2g^{2}\beta^{4}\gamma_{R})$ and carrying out some derivations
we obtain,
\begin{eqnarray}
&& F_{R}[u,s] = -a\,u^4 + {\cal U}_{R}[u,s]\,, \\
&& {\cal U}_{R}[u,s] = {\cal U}_{R}[s] + {\cal V}_{R}[u,s]\,, \\
&& {\cal V}_{R}[u,s] = \frac{\beta}{2\pi R}\sum_{n=-\infty}^{\infty}
\log \frac{L_{R}((n+s)^{2}+u^{2})\,L_{R}((n+s)^{2}-u^{2}) }
{L_{R}^{2}((n+s)^{2})}\,, \\
&& {\cal U}_{R}[s] = \frac{\beta}{\pi R}\sum_{m=0}^{\infty}\log \left( 1 -
\frac{\cos 2\pi s}{\cosh (\pi(m+1/2)\beta/R)}\right)\,,
\end{eqnarray}
where $L_{R}(x) = \cosh (2\pi R\sqrt{x}/\beta)$.

It can be seen that
at finite $R$ this free energy possesses only a trivial minimum at $s=u=0$.
The situation changes after taking the thermodynamic limit, 
$R\rightarrow \infty$. The resulting expression for the free energy
density (see Ref. \cite{SveTim}) possesses only the trivial stable solution
$u=s=0$ at high temperatures. However, at some critical
temperature $T_c$ the system undergoes a first order phase
transition, below which there appears a deeper nontrivial minimum 
at $u = s =1/2$ (see Ref. \cite{SveTim}).

Therefore, at high temperatures $T > T_{c}$ the dependence $Z[\chi]$ is 
determined by the solution of the linearised equation 
$\delta^2W/\delta \varsigma^2 \bullet \sigma_1$ around $\varsigma=0$.
This can only produce the dependence akin to the Abelian theory. 
Namely, it contains the delta--function expressing the conservation of
the global charge (Eq. (\ref{Zzchi}) with $\rho=0$), 
and apart from that it is trivial 
in the sense that $Z[\chi_{lm}]\rightarrow 1$ in the thermodynamic limit 
$R\rightarrow\infty$.
This situation, obviously, corresponds to the {\it deconfinement} phase
as there is no restriction on the colour fluxes at infinity.

However, below the critical temperature $T< T_c$
there is a nonzero constant solution
$|\varsigma|=\pi/g\beta$. Since the system in invariant under
the group of the {\it big gauge transformations}
$G_\infty$ parametrised by matrices
$U(\hx)$, the corresponding unit colour vector $\hat{\bbox{\varsigma}}(\hx)$
is arbitrary in every direction $\hx$.
After integration over the orbits of the group SU(2) at each cone $\hx$
the dependence becomes,
\begin{eqnarray}
\tilde{Z}[\chi] &=& \prod_{\hx} \frac{\sin \Delta\beta\,\varsigma \chi(\hx)}
{\Delta\beta\,\varsigma \chi(\hx)} 
\sim  \exp\left(-\frac{\Delta\pi^2}{g^2}
\int_{\partial V} d\hx\, \chi^2(\hx)\right)\, 
\end{eqnarray}
where we have introduced a discretisation of the unit sphere with 
$\Delta$ being the infinitesimal cone area.
It is well known \cite{FadSlav} that in the continuous limit the bare 
coupling constant vanishes $g\rightarrow 0$ thereby
making the effective partition function $Z[\chi]$ a very sharply 
peaked function around the zero argument due to its essentially
non--perturbative dependence on $g$. 
This property corresponds to the {\it confinement}
phase, in which colour fluxes are equal to zero in every spatial direction
at infinity.

So, we can conclude that the dependence of the effective
free energy on $\chi$ is the following,
\begin{equation}
Z[\chi] = \left\{ \begin{array}{ll}
\prod_{\hx}\delta(\chi(\hx)), & T < T_{c},\\
1, & T>T_{c}. \\ 
\end{array}\right. 
\end{equation}
Therefore, the Gibbs average of an observable $A$ is given by,
\begin{equation} \label{AVv}
<A> = \left\{ \begin{array}{ll}
\langle A \rangle_{0}, & T < T_{c},\\
\int d\chi(\hx)\, \langle A \rangle_{\chi}, & T>T_{c}, \\ 
\end{array}\right. 
\end{equation}
where we have introduced the averages over inequivalent representations,
\begin{equation}
\langle A \rangle_{\chi} = \frac{1}{Z[\chi]}
\int {\cal D}\sigma\,{\cal D}\nu\, e^{-W[\sigma,\nu]
+ i\int_{\partial\Lambda} dt\,d\hx\,\sigma\chi}\,A[\sigma,\nu]\,.
\end{equation}
It is straightforward to see that any Gibbs average at
low temperatures contains the singlet projector of the group $G_\infty$,
\begin{equation}
\langle A \rangle _{0} = 
\lim_{R\rightarrow\infty}\frac{1}{Z_{R}[0]} \mbox{Tr}\, 
(e^{-\beta H_{R}}\, \delta(Q_R)\, A) =
\lim_{R\rightarrow\infty} \frac{1}{Z_{R}[0]}\mbox{Tr}\, 
(e^{-\beta H_{R}}
\,P_{s}\,A)\,,
\end{equation}
where $Q_R=\int_{V_R} d\bbox{x}\, \bbox{\nabla E}$ is the operator of 
the colour charge in volume $V$
and $P_{s}$ is the singlet projector of the big gauge transformations.
Presence of this projector in the Gibbs averages has been
demonstrated to lead to the area law for
the Wilson loop \cite{Wils,SveTim}, 
what is considered to be a standard confinement criterion.

%Concl
\section{Conclusion}

The Gauss law in the gauge field theory may be resolved explicitly
in a physical gauge. This produces
effectively non--local interactions generating a boundary nontriviality
of the theory.

We have studied  the dependence of the effective partition function 
on the Dirichlet boundary condition $R^{2}E_{\Vert}(R\hat{\bbox{x}}) = 
\chi(\hat{\bbox{x}})$ imposed on the residual 
component of the electric field for electrodynamics 
with an external charge and SU(2) gluodynamics.
In the Abelian case this dependence always contains the delta function
expressing the conservation of the total charge, and it is
nontrivial only for charge distributions
slowly decreasing at spatial infinity.

The non--Abelian self--interactions lead to a more unusual effect.
Here the restriction of possible boundary values of 
the longitudinal electric field at low temperatures
provides the confinement mechanism proposed by us in  Ref. \cite{SveTim}.
Indeed, this quantity is proportional to the flux
of the electric field through an infinitesimal cone in the direction
$\hat{\bbox{x}}$ at infinity. 
Therefore, since the colour flux vanishes for any direction, 
no colour could escape to infinity and be experimentally observed.

\acknowledgments

The authors are grateful to Professors B.A.~Arbuzov, V.I.~Savrin,
Drs. E.E.~Boos, V.A.~Ilyin, V.O.~Soloviev, 
and Yu.A.~Kuznetsov for numerous fruitful discussions.

\appendix

\section{Generalised Fock--Schwinger gauge}

Let $V_{R}$ be a regular domain in ¢ $\bbox{R}^{3}$ topologically
equivalent to a ball with a smooth boundary $\partial V_{R}$. 
One can choose the curvilinear coordinate system
$\BX$ in $V_{R}$ such that on the boundary
$\partial V_{R}$ the first coordinate is constant and equal to
some parameter $R$, which would play the role of an infrared
regulator, i.e.
$\partial V_{R} =\{\BX: X_{1}(\Bx) = R =const\}$.
The field of vectors normal to the boundary for all possible
values of $R$ forms a differentiable vector field in $\bbox{R}^3$.
The ordinary Cartesian coordinates  we shall denote as $\Bx$ to distinguish
them from $\BX$.
The local orthonormal frame then can be written as,
\begin{equation} \label{G-3}
e^{(k)}_{i} = \frac{1}{h_{k}}\,\frac{\partial x_{i}}{\partial X_{k}}\,,
 \quad h_{k} = \left( \sum_{i = 1}^{3} \left( \frac{\partial x_{i}}
{\partial X_{k}} \right)^{2} \right)^{1/2}, 
\quad h \equiv \prod_{i=1}^{3}h_{i},
\end{equation}
where $h_k$ are called the Lam\'e coefficients.
Also, to distinguish the components of vectors in the curvilinear
frame we shall use the notations with indices in parenthesis,
\begin{equation} \label{G-4}
A_{(k)} = e^{i}_{(k)}\,A_{i}, \quad \partial_{(k)} \equiv \frac{1}{h_{k}}\,
\frac{\partial}{\partial X_{k}}.
\end{equation}
Then $\bbox{e}_{(1)}$ defines the field of normal vectors
we have just introduced. It is natural to introduce
the (2+1) decomposition onto the longitudinal and transversal
(denoted by the Greek characters) components:
$i \rightarrow (1, \alpha ),\ \alpha = 2,3$, where
the corresponding 2-d radius--vector will be denoted as $\BrX = (X_{2},X_{3})$.

The gauge theory in a finite domain $V_{R}$ acquires especially elegant
formulation in a gauge which is consistent with the shape of the boundary.
Namely, we shall require that the normal component of the gauge 
field vanishes in every point,
\begin{eqnarray} \label{G-5}
\bbox{e}_{(1)}(\Bx)\,\bbox{A}(t,\Bx) = 0,\ \, \bbox{A}=\bbox{A}_{\perp},
\quad \bbox{A}_{\perp}= P \bbox{A},\ \, P =\bbox{1} -\bbox{e}_{(1)}
\bigotimes \bbox{e}_{(1)}.
\end{eqnarray}
This gauge condition is natural to name the {\em generalised
Fock--Schwinger gauge}.
Its most significant property is that the Gauss law,
$\nabla_{i} E_{i}=0$,
can be resolved explicitly,
\begin{equation}
\label{G-6}
E_{(1)} = - \frac{h_{1}}{h} \int_{X^{(0)}_{1}}^{X_{1}}
dX_{1}'\, (h\Phi_{\perp})(X_{1}',\BrX),
\end{equation}
expressing the longitudinal component of the electric field
through the transversal components of the gauge and strength fields,
\begin{equation} \label{G-6p}
\Phi_{\perp} \equiv \nabla_{i} E_{\perp\,i} =
\frac{h_{\alpha}}{h}\nabla_{(\alpha)}\left(
\frac{h}{h_{\alpha}}E_{(\alpha)}\right).
\end{equation}
Above the lower integration limit
$X^{(0)}_{1}$ is equal to some constant, which should be chosen
in the reference point of our coordinate system.
Analogously, one can solve the identity,
\begin{equation}\label{G-7}
\nabla_{i}G_{i} = 0, \quad G_{i} \equiv \nabla_{j} F_{ij},
\end{equation}
expressing $G_{(1)}$ as,
\begin{equation} \label{G-8}
G_{(1)} = - \frac{h_{1}}{h} \int_{\bar{X}^{(0)}_{1}}^{X_{1}}
dX_{1}'\, (h\ \nabla_{i} G_{\perp\,i})(X_{1}',\BrX).
\end{equation}
Further, one of the components of the Bianchi identity is,
\begin{equation} \label{G-9}
\nabla_{i} B_{i} =0, \quad B_{k} = \frac{1}{2}\epsilon_{ijk}F_{ij},
\end{equation}
what allows us to express the longitudinal magnetic field as,
\begin{equation} \label{G-10}
B_{(1)} = - \frac{h_{1}}{h} \int_{X^{(0)}_{1}}^{X_{1}}
dX_{1}'\, (h\ \nabla_{i} B_{\perp\,i})(X_{1}',\BrX).
\end{equation}
Other components of the Bianchi identity
$\epsilon_{ijk}\,e^{i}_{(\alpha)} \nabla_{j} E_{k} = 0$
give,
\begin{equation} \label{G-12}
\frac{1}{h_{\alpha}}\frac{\partial}{\partial X_{1}} \left(
h_{\alpha} E_{(\alpha)} \right) = \nabla_{(\alpha)} \left(
h_{1} E_{(1)}\right),
\end{equation}

A nice property of the Fock--Schwinger gauge is that the
gauge strength and potentials are related to each other in a simple way.
Indeed, from the definition of the gauge strength tensor applying the
gauge condition we have,
\begin{equation} \label{G-13}
\frac{1}{h_{\alpha}}\frac{\partial}{\partial X_{1}} \left(
h_{\alpha} A_{(\alpha)} \right) = h_{1} F_{(\alpha)(1)}.
\end{equation}
However, the integral form of this equation, 
\begin{equation} \label{G-14}
A_{(\alpha)}=\frac{1}{h_{\alpha}}\int_{\tilde{X}_{1}^{(0)}}^{X_{1}}
dX_{1}'\,(h_{1}h_{\alpha}F_{(\alpha)(1)})(X_{1}',\BrX),
\end{equation}
breaks the residual gauge transformations symmetry subgroup,
whereas choosing the lower integration limits in all
previous integral relations did not violate such a symmetry. 
Fixing some kind of the boundary condition above,
e.g. choosing $\tilde{X}_{1}^{(0)}=x^{0}_{1}$ 
in the reference point, so that
\begin{equation} \label{G-15}
\lim_{X_{1}\rightarrow x_{1}^{0}}(h_{\alpha}A_{(\alpha)})(\bbox{X}) = 0,
\end{equation}
can be shown to be sufficient for determining
a unique gauge field satisfying the gauge condition.
Indeed, let $\tilde{\bbox{A}}$ be any gauge field.
The transformation to the Fock--Schwinger gauge is accomplished
by a gauge transformation $U(\bbox{X})$,
\begin{equation}\label{G-16}
\bbox{A}(\bbox{X}) = U^{-1}\,(\,\tilde{\bbox{A}}(\bbox{X})-g^{-1}\,\partial \,)
\,U(\bbox{X})\,, 
\quad \bbox{E}(\bbox{X}) = U^{-1}\tilde{\bbox{E}}(\bbox{X})\,U(\bbox{X}),
\end{equation}
which can be found from the equation,
\begin{equation}\label{G-17}
\frac{1}{h_{1}}\frac{\partial }{\partial \,X_{1}}U(\bbox{X}) =
 g (\bbox{e}_{(1)}\,\tilde{\bbox{A}})(\bbox{X})\,U(\bbox{X})\,.
\end{equation}
For the moment we can choose the initial condition simply as
$U(X_{1}=0,\BrX)=1$.
The solution of (\ref{G-17}) is given by the Dyson P-exponent,
\begin{equation}\label{G-18}
U(\bbox{X}) = P\exp\int_{0}^{1}d\alpha\,R(\alpha, \bbox{X}),
\ \, R(\alpha,\bbox{X}) = g X_{1}(h_{1}\bbox{e}_{(1)}\tilde{\bbox{A}})
(\alpha X_{1},\BrX).
\end{equation}
This can be explicitly worked out for the gauge fields,
\begin{eqnarray}
&&\bbox{A}^{b}(\bbox{X}) =\tilde{\bbox{A}}^{a}({\bf
X})\,P\exp\int _{0}^{1}d\alpha \,(-gt^{abc}\,R^{c}(\alpha,\bbox{X})\,)- \nonumber\\
 &&- g^{-1}\int _{0}^{1}d\beta \,\partial R^{a}(\beta ,\bbox{X})\,P\exp\int _{0}^{\beta }d\gamma \,
(-gt^{abc}\,R^{c}(\gamma ,\bbox{X})\,).\,
\label{G-19}
\end{eqnarray}
By applying an additional residual gauge transformation we can always
satisfy the boundary condition (\ref{G-15}), and uniquely.
Really, suppose there exist two distinct gauge fields
$\bbox{A}'$ and $\bbox{A}''$ satisfying Eqs. (\ref{G-5},\ref{G-15})
and such that
$\bbox{ A}'' \neq  U^{-1} \bbox{ A}' U$, $ \forall \ U = \mbox{const}$.
Then there should exist a gauge transformation between the two,
\begin{equation} \label{G-20}
\bbox{A}''(\bbox{X})  =  U^{-1}(\bbox{X})\  (\bbox{ A}'({\bf
X})-g^{-1}\,\partial \,)\ U(\bbox{X})\,.
\end{equation}
Multiplication of this by $\bbox{e}_{(1)}$ gives
$(1/h_{1})(\partial \,U(\bbox{X})/\partial \,X_{1}) =  0\,$,
i.e. $U = U(\BrX)$.  Such transformations form the subgroup 
$G_{res}$ of the residual gauge symmetries in the Fock--Schwinger gauge.
In fact, the boundary condition
(\ref{G-15}) does not permit such transformation since in the limit
$\bbox{X} \rightarrow \bbox{x}^{0}$ in Eq. 
(\ref{G-20}) the gauge fields would have the singularity which is
not compatible with such a boundary condition. 
This proves the uniqueness of the gauge field.

As a simple example let us consider the {\em elliptic coordinates}:
$1 \le X_{1} <\infty$, $-1 \le X_{2} \le 1$, $0 \le X_{3} \le 2\pi$,
where $X_{3}=\phi$ is the polar angle and $X_{1} =(r_{1}+r_{2})/2a$,
$X_{2}=(r_{1}-r_{2})/2a$. 
Such a coordinate system is defined by two points located at the distances
$\pm a$ from the reference point along the $z$ axis
with $\bbox{r}_{1}$, $\bbox{r}_{2}$ being the radius--vectors from these points. 
The reference point is here  $x^{0}_{1} = 1$,
$x^{0}_{2} = 0$ and the Lam\'e coefficients are given by,
\begin{equation}\label{G-21}
h_{1}^{2} =a^{2}\frac{X_{1}^{2}-X_{2}^{2}}{X_{1}^{2}-1},\ 
h_{2}^{2} =a^{2}\frac{X_{1}^{2}-X_{2}^{2}}{1-X_{2}^{2}},\ 
h_{3}^{2} =a^{2}(X_{1}^{2}-1)(1-X_{2}^{2}).
\end{equation}
In the limit $a=0$ 
the ellipsoid becomes a sphere and the gauge turns into the 
standard Fock--Schwinger gauge. For the {\em spherical coordinates}
$X_{1}=r$, $X_{2}=\phi$, $X_{3}=\theta$ 
the Lam\'e coefficients become very simple,
\begin{equation}\label{G-21p}
h_{1}=1,\quad h_{2}=X_{1}\sin X_{3},\quad h_{3} = X_{1}. 
\end{equation}
A technically attractive property of this particular gauge is that
$\bbox{e}_{(i)}$ do not depend on $X_{1}$
and the vector normal to the boundary is just equal
to the unit radius--vector $\bbox{e}_{(1)}=\Hx$.

%Bibl

\end{document}